\documentclass[aps,twocolumn,prl]{revtex4}
\usepackage{amsfonts}
\usepackage[dvips]{graphicx}
\usepackage[T1]{fontenc}
\usepackage{amsmath,amsbsy,amssymb,graphics}
\def\beginABC{\begin{subequations}}
\def\endABC{\end{subequations}}
\let\mathbf=\boldsymbol

\begin{document}

\title{Supersymmetry and Unconventional Quantum Hall Effect in Graphene}
\author{Motohiko Ezawa}
\affiliation{Department of Physics, University of Tokyo, Hongo 7-3-1, 113-0033, Japan }
\date{\today}

\begin{abstract}
We present a unified description of the quantum Hall effect in graphene on
the basis of the 8-component Dirac Hamiltonian and the supersymmetric (SUSY)
quantum mechanics. It is remarkable that the zero-energy state emerges
because the Zeeman splitting is exactly as large as the Landau level
separation, as implies that the SUSY is a good symmetry. For nonzero energy
states, the up-spin state and the down-spin state form a supermultiplet
possessing the spin SU(2) symmetry. We extend the Dirac Hamiltonian to
include two indices $j_{\uparrow }$ and $j_{\downarrow }$, characterized by
the dispersion relation $E\left( p\right) \propto p^{j_{\uparrow
}+j_{\downarrow }}$ and the Berry phase $\pi (j_{\uparrow }-j_{\downarrow })$%
. The quantized Hall conductivity is shown to be $\sigma _{xy}=\pm \left(
2n+j_{\uparrow }+j_{\downarrow }\right) 2e^{2}/h$.
\end{abstract}

\maketitle

The quantum Hall effect (QHE) is one of the most remarkable phenomena in
condensed matter discovered in the last century\cite{BookDasSarma,BookEzawa}%
. Electrons, undergoing cyclotron motion in magnetic field, fill Landau
levels successively. Each filled energy level contributes one conductance
quantum $e^{2}/\hbar $ to the Hall conductivity $\sigma _{xy}$. The Hall
plateau develops at $\sigma _{xy}=\nu (e^{2}/h)$, where $\nu $ is the
filling factor. It tells us how many energy levels are filled up. Hall
plateaux have been observed at $\nu =1,2,3,\ldots $ in the conventional
semiconductor QHE.

\begin{figure}[h]
\includegraphics[width=0.46\textwidth]{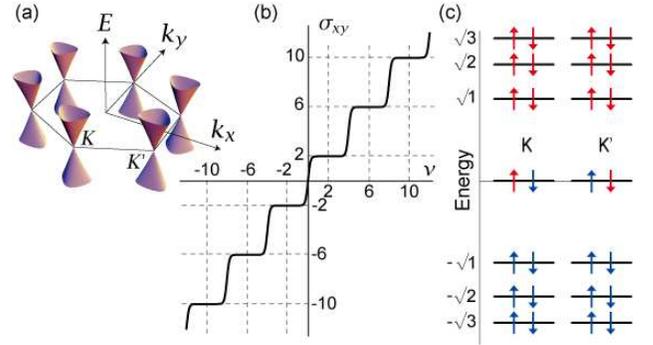}
\caption{{}(a) A schematic diagram of the low-energy dispersion relation
near the Dirac points (K and K' points) in the graphene Brillouin zone. Only
two Dirac cones are inequivalent to each other, producing a 2-fold valley
degeneracy in the band structure. (b) The quantized Hall conductivity as a
function of the filling factor $\protect\nu $ in monolayer graphene. (c) The
Landau level spectrum of monolayer graphene for electrons (red) and holes
(blue) at the K and K' points. The arrow indicates the spin. This spectrum
is a manifestation of SUSY.}
\label{FigGraphQHE}
\end{figure}

Recent experimental developments have revealed unconventional QHE in graphene%
\cite{Nov1,Nov2,Zhang,Nov3}. The filling factors\cite{Ando,Gusynin,Peres}
form a series [Fig.\ref{FigGraphQHE}(b)], 
\begin{equation}
\nu =\pm 2,\pm 6,\pm 10,\cdots ,
\end{equation}
where the basic height in the Hall conductance step is $4e^{2}/h$ except for
the first step which is just one half. In the bilayer graphene QHE, the
series reads\cite{Nov3,MacCann}, 
\begin{equation}
\nu =\pm 4,\pm 8,\pm 12,\cdots ,
\end{equation}
where all steps have the same height $4e^{2}/h$. A most recent experiment%
\cite{Zhang06L} has shown a fine structure at $\nu =0,\pm 1,\pm 4$ in
monolayer graphene.

In this paper, we present a unified description of the QHE in graphene on
the basis of the 8-component Dirac Hamiltonian and the supersymmetric (SUSY)
quantum mechanics\cite{Witten}. Our first observation is that the Zeeman
splitting is as large as the Landau-level separation, as leads to the
emergence of the zero energy state. Furthermore, the spin SU(2) symmetry is
exact within each Landau level for nonzero energy states. Our second
observation is that the above Landau-level structure is a manifestation of
SUSY at the K and K' points. 

We first formulate the QHE in monolayer graphene. Then, in order to make the
underlying mathematical structure clear, we generalize our analysis to the
system where the dispersion relation has the form $E\left( p\right) \propto
p^{j_{\uparrow }+j_{\downarrow }}$ and the Berry phase is given by $\pi
(j_{\uparrow }-j_{\downarrow })$ in the absence of the magnetic field. The
monolayer graphene corresponds to $j_{\uparrow }=1$ and $j_{\downarrow }=0$,
while the bilayer graphene to $j_{\uparrow }=2$ and $j_{\downarrow }=0$. We
show that the zero-energy states have the $j_{\uparrow }$-fold ($%
j_{\downarrow }$-fold) degeneracy for up-spin (down-spin) electrons at the K
point. As a result the quantized values of the Hall conductivity become%
\begin{equation}
\sigma _{xy}=\pm \left( n+\frac{j_{\uparrow }+j_{\downarrow }}{2}\right) 
\frac{4e^{2}}{h},\qquad n=0,1,2,\cdots .  \label{HallConduA}
\end{equation}%
We also discuss the relation of our SUSY formalism to the Atiyah-Singer
index theorem.

We\ start with the SUSY description of the monolayer graphene. The
low-energy band structure of graphene can be approximated as cones located
at two inequivalent Brillouin zone corners called the K and K' points [Fig.%
\ref{FigGraphQHE}(a)]. In these cones, the two-dimensional energy dispersion
relation is linear and the electron dynamics can be treated as
`relativistic' Dirac electrons\cite{Ajiki}, in which the Fermi velocity $v_{%
\text{F}}$ of the graphene substitutes for the speed of light.

Corresponding to the K point ($+$) and the K' point ($-$), we have two Dirac
Hamiltonians%
\begin{equation}
H_{\text{D}}^{\pm }=v_{\text{F}}(\alpha _{x}P_{x}\pm \alpha _{y}P_{y})+\beta
mv_{\text{F}}^{2},  \label{DiracHamilD}
\end{equation}%
where $P_{i}\equiv -i\hbar \partial _{i}+eA_{i}$ is the covariant momentum,
and%
\begin{equation}
\alpha _{i}=\left( 
\begin{array}{cc}
0 & \sigma _{i} \\ 
\sigma _{i} & 0%
\end{array}%
\right) ,\quad \beta =\left( 
\begin{array}{cc}
\mathbb{I}_{2} & 0 \\ 
0 & -\mathbb{I}_{2}%
\end{array}%
\right)
\end{equation}%
with $\sigma ^{i}$ the Pauli matrix for the spin degree of freedom and $%
\mathbb{I}_{2}$ the $2\times 2$ unit matrix. We have introduced the vector
potential $A_{i}$, in terms of which the magnetic field is $\mathbf{B}=%
\mathbf{\nabla }\times \mathbf{A}$. We assume a homogeneous magnetic field $%
\mathbf{B}=(0,0,-B)$ with $B>0$. In the zero field case, the Hamiltonian (%
\ref{DiracHamilD}) has a linear dispersion $E(p)=\pm v_{F}p$ for both spin
states and for both K and K' points. The mass of quasiparticle excitations
is zero, $m=0$, in the naive band-structure calculation for noninteracting
quasiparticles on the hexagonal lattice of graphene. However, it is natural
to expect a nonzero excitation gap when the Coulomb interaction is taken
into account. Hence we have included the effective mass term into the Dirac
Hamiltonian (\ref{DiracHamilD}). Our analysis is valid whether $m=0$ or $%
m\neq 0$.

The Hamiltonian (\ref{DiracHamilD}) is expressed as\cite{Thaller92} 
\begin{equation}
H_{\text{D}}^{\pm }=\left( 
\begin{array}{cc}
mv_{\text{F}}^{2} & Q_{\pm } \\ 
Q_{\pm } & -mv_{\text{F}}^{2}%
\end{array}%
\right) ,  \label{DiracHamilB}
\end{equation}%
with 
\begin{equation}
Q_{\pm }=v_{\text{F}}\left( \sigma _{x}P_{x}\pm \sigma _{y}P_{y}\right) .
\end{equation}%
It is diagonalized,%
\begin{equation}
H_{\text{D}}^{\pm }=\text{diag.}\left( \sqrt{Q_{\pm }Q_{\pm }+m^{2}v_{\text{F%
}}^{4}},-\sqrt{Q_{\pm }Q_{\pm }+m^{2}v_{\text{F}}^{4}}\right) ,
\label{DiracHamilC}
\end{equation}%
where the negative component describes holes.

It is convenient to distinguish electrons on the K and K' points by
assigning the pseudospin to them. Namely, we call the electron on the K (K')
point the up-pseudospin (down-pseudospin) electron. We can combining them
into the 8-component Dirac Hamiltonian,%
\begin{equation}
H_{\text{D}}=v_{\text{F}}\alpha _{x}P_{x}+v_{\text{F}}\tau _{z}\alpha
_{y}P_{y}+\beta mv_{\text{F}}^{2}=\text{diag.}(H_{\text{D}}^{+},H_{\text{D}%
}^{-}),  \label{DiracHamilA}
\end{equation}%
where $\tau _{z}$ is the Pauli matrix acting on the pseudospin space. The
pseudospin is reversed by the operation 
\begin{equation}
\tau _{x}H_{\text{D}}\tau _{x}=\text{diag.}(H_{\text{D}}^{-},H_{\text{D}%
}^{+}).
\end{equation}
Furthermore, the Hamiltonian $H_{\text{D}}^{\pm }$ has a symmetry, 
\begin{equation}
\Gamma H_{\text{D}}^{\pm }\Gamma =-H_{\text{D}}^{\pm },
\end{equation}
with $\Gamma =$diag.$(\sigma _{z},\sigma _{z})$. It implies that a solution $%
|\Psi \rangle $ to the Dirac equation with energy $E$ has a particle-hole
conjugate partner $\Gamma |\Psi \rangle $ with energy $-E$. Therefore, it is
enough to explore the energy spectrum of electrons for $E\geq 0$ at the K
point. Nevertheless, to make the underlying physical and mathematical
structure clear, we present all the spectrum in what follows.

We consider the quantity\cite{Thaller92} 
\begin{equation}
H^{\pm }=2Q_{\pm }Q_{\pm }=2v_{\text{F}}^{2}\left( -i\hbar \nabla +e\mathbf{A%
}\right) ^{2}\mp 2e\hbar v_{\text{F}}^{2}\sigma _{z}B.  \label{PauliHamil}
\end{equation}%
Since this has the same form as the Pauli Hamiltonian with the mass $m^{\ast
}=1/4v_{\text{F}}^{2}$ except for the dimension, we call it the Pauli
Hamiltonian for brevity. The energy spectrum $\mathcal{E}_{n}$ of the Dirac
Hamiltonian is constructed once we know the one $E_{n}$ of the Pauli
Hamiltonian (\ref{PauliHamil}). It is to be noticed in (\ref{PauliHamil})
that the direction of the magnetic field is effectively opposite at the K
and K' points.

We introduce a pair of operators 
\begin{equation}
a=\frac{l_{B}}{\sqrt{2}\hbar }(P_{x}+iP_{y}),\qquad a^{\dagger }=\frac{l_{B}%
}{\sqrt{2}\hbar }(P_{x}-iP_{y})
\end{equation}
with the magnetic length $l_{\text{B}}=\sqrt{\hbar /eB}$. The commutation
relation $\left[ a,a^{\dagger }\right] =1$ follows from $[P_{x},P_{y}]=i%
\hbar ^{2}/l_{B}^{2}$. The operator $Q_{\pm }$ is expressed as 
\begin{equation}
Q_{+}=\left( 
\begin{array}{cc}
0 & A^{\dagger } \\ 
A & 0%
\end{array}%
\right) ,\qquad Q_{-}=\left( 
\begin{array}{cc}
0 & A \\ 
A^{\dagger } & 0%
\end{array}%
\right) .  \label{SuperQ}
\end{equation}%
with $A=\hbar \omega _{c}a$, where $\omega _{c}=\sqrt{2}v_{\text{F}%
}/l_{B}=v_{\text{F}}\sqrt{2eB/\hbar }$.

\begin{figure}[h]
\begin{center}
\includegraphics[width=0.47\textwidth]{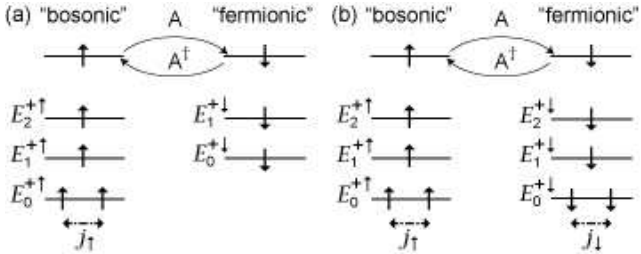}
\end{center}
\caption{{}The energy spectrum of the superpartner Hamiltonians $%
H^{+\uparrow }$ and $H^{+\downarrow }$ for electrons at the K point. Two
states on the same horizontal line have the same energy, making a
supermultiplet, except for the zero-energy state ($E_{0}^{+\uparrow }=0$).
The zero-energy state is $j_{\uparrow }$-fold ( $j_{\downarrow }$-fold)
degenerated for up-spin (down-spin) electrons. (a) The case with $%
j_{\downarrow }=0$. We have $j_{\uparrow }=1$ in monolayer graphite, while $%
j_{\uparrow }=2$ in bilayer graphite. (b) The case with $j_{\downarrow }\neq
0$.}
\label{FigSuperPair}
\end{figure}

Let us first make a generic argument that is valid for an arbitrary operator 
$A$. We focus on the Hamiltonian $H^{+}=$diag.$(H^{+\uparrow
},H^{+\downarrow })$ related to the K point, with $H^{+\uparrow }=A^{\dagger
}A$ and $H^{+\downarrow }=AA^{\dagger }$. This is a simplest example of the
SUSY quantum mechanics\cite{Witten}, where the superalgebra reads 
\begin{equation}
H^{+}=\{Q_{+},Q_{+}\},\qquad \left[ H^{+},Q_{+}\right] =0,
\end{equation}
with $Q_{+}$ the supercharge. The two Hamiltonians $H^{+\uparrow }$ and $%
H^{+\downarrow }$ are superpartners. We consider separately the eigenvalue
problems for the up-spin and down-spin components, 
\begin{equation}
H^{+\uparrow \downarrow }|\psi _{n}^{+\uparrow \downarrow }\rangle
=E_{n}^{+\uparrow \downarrow }|\psi _{n}^{+\uparrow \downarrow }\rangle ,
\end{equation}%
where $E_{n+1}^{+\uparrow \downarrow }>E_{n}^{+\uparrow \downarrow }\geq
E_{0}^{+\uparrow \downarrow }$. We assume $E_{0}^{+\uparrow }=0$, as implies
that the SUSY is a good symmetry\cite{Witten}. (This is indeed the case in
our system with $A=\hbar \omega _{c}a$.) Using the relations 
\begin{equation}
AH^{+\uparrow }=AA^{\dagger }A=H^{+\downarrow }A,
\end{equation}%
we obtain\beginABC\label{SuperSerieA}%
\begin{align}
H^{+\downarrow }A|\psi _{n}^{+\uparrow }\rangle & =AH^{+\uparrow }|\psi
_{n}^{+\uparrow }\rangle =E_{n}^{+\uparrow }A|\psi _{n}^{+\uparrow }\rangle ,
\\
H^{+\uparrow }A^{\dagger }|\psi _{n}^{+\downarrow }\rangle & =A^{\dagger
}H^{+\downarrow }|\psi _{n}^{+\downarrow }\rangle =E_{n}^{+\downarrow
}A^{\dagger }|\psi _{n}^{+\downarrow }\rangle .
\end{align}%
\endABC If $E_{n}^{+\uparrow }\neq 0$, $A|\psi _{n}^{+\uparrow }\rangle $ is
an eigenstate of $H^{+\downarrow }$. Similarly, if $E_{n}^{+\downarrow }\neq
0$, $A^{\dag }|\psi _{n}^{+\downarrow }\rangle $ is an eigenstate of $%
H^{+\uparrow }$. Thus, there is one-to-one correspondence between the
up-spin eigenstate and the down-spin eigenstate for nonzero energy states
[Fig.\ref{FigSuperPair}]. They are said to make a supermultiplet, since the
correspondence is made by the supercharge $Q_{+}$. We identify\cite{Witten}
the up-spin (down-spin) sector as the bosonic (fermionic) sector at the K
point.

It is necessary to examine the cases $E_{0}^{+\downarrow }\neq 0$ and $%
E_{0}^{+\downarrow }=0$, separately. If $E_{0}^{+\downarrow }\neq 0$, the
lowest energy eigenstate of $H^{+\downarrow }$ is $|\psi _{0}^{+\downarrow
}\rangle \propto A|\psi _{1}^{+\uparrow }\rangle $. Then, the states $|\psi
_{n+1}^{+\uparrow }\rangle $\ and $|\psi _{n}^{+\downarrow }\rangle $ make a
supermultiplet having the same energy $E_{n+1}^{+\uparrow
}=E_{n}^{+\downarrow }$. Namely, there hold the relations [Fig.\ref%
{FigSuperPair}(a)]\beginABC%
\begin{equation}
|\psi _{n}^{+\downarrow }\rangle =\frac{1}{\sqrt{E_{n+1}^{+\uparrow }}}%
A|\psi _{n+1}^{+\uparrow }\rangle ,\quad |\psi _{n+1}^{+\uparrow }\rangle =%
\frac{1}{\sqrt{E_{n}^{+\downarrow }}}A^{\dag }|\psi _{n}^{+\downarrow
}\rangle  \label{SuperMultiA}
\end{equation}%
for $n\geq 0$. On the other hand, if $E_{0}^{+\downarrow }=0$, for which $%
A^{\dag }|\psi _{0}^{+\uparrow }\rangle =0$ additionally, there hold the
relations [Fig.\ref{FigSuperPair}(b)]%
\begin{equation}
|\psi _{n}^{+\downarrow }\rangle =\frac{1}{\sqrt{E_{n}^{+\uparrow }}}A|\psi
_{n}^{+\uparrow }\rangle ,\quad |\psi _{n}^{+\uparrow }\rangle =\frac{1}{%
\sqrt{E_{n}^{+\downarrow }}}A^{\dag }|\psi _{n}^{+\downarrow }\rangle
\end{equation}%
\endABC for $n\geq 1$. The similar analysis is applicable also to the
Hamiltonian related to the K' point, where $H^{-}=$diag.$(H^{-\uparrow
},H^{-\downarrow })$ with $H^{-\downarrow }=A^{\dagger }A$ and $H^{-\uparrow
}=AA^{\dagger }$. It is to be noticed that the bosonic (fermionic) sector is
identified with the down-spin (up-spin) states at the K' point. The energy
spectrum is 4-fold degenerated except for the zero-energy state. We cannot
say anything about the degeneracy of the zero-energy state by this general
argument.

We apply the above analysis to the monolayer graphene with the "Dirac"
Hamiltonian (\ref{DiracHamilC}) with (\ref{SuperQ}), where $A=\hbar \omega
_{c}a$ and $E_{0}^{+\uparrow }=0$. Using the commutation relation $\left[
a,a^{\dagger }\right] =1$, it is trivial to see that the energy eigenvalues
of the Dirac Hamiltonian $H_{\text{D}}^{\pm }$ are 
\begin{equation}
\mathcal{E}_{0}^{+\uparrow }=\mathcal{E}_{0}^{-\downarrow }=\pm mv_{\text{F}%
}^{2}
\end{equation}%
and 
\begin{align}
\mathcal{E}_{n+1}^{+\uparrow } &=\mathcal{E}_{n}^{+\downarrow }=\mathcal{E}%
_{n+1}^{-\downarrow }=\mathcal{E}_{n}^{-\uparrow }  \notag \\
&=\pm \hbar \omega _{c}\sqrt{n+1+(m^{2}v_{\text{F}}^{4}/\hbar ^{2}\omega
_{c}^{2})}
\end{align}%
for $n\geq 0$. If $m=0$, there exists one zero-energy state for up-spin
electrons but not for down-spin electrons at the K point [Fig.\ref%
{FigSuperPair}(a)]. The existence of the zero energy state is an intriguing
property of the SUSY\ theory, where the bosonic and fermionic zero-point
energies are canceled out\cite{Witten}. The physical reason is that the
Zeeman splitting is exactly as large as the Landau level separation. This is
a well known property\cite{Thaller92} of the Dirac electron in magnetic
field though it is overlooked in all previous literature on the QHE in
graphene. On the other hand, there exists one zero-energy state for
down-spin electrons but not for up-spin electrons at the K' point. Recall
that the direction of the magnetic field is effectively opposite at the K
and K' points.

There exists the $4$-fold degeneracy in the zero-energy state due to
electrons and holes. No plateau is made at $\nu =0$ because the Landau level
made of the zero energy states is half-filled. However, the degeneracy is
removed between electrons and holes once the mass term is present. However
small the mass $m$ may be, the Hall plateau emerges at $\nu =0$ since holes
are filled before electrons. Thus, by assuming $m\neq 0$, we can explain a
recent experimental data\cite{Zhang06L} at $\nu =0$.

We next investigate a general case, where $A$ is given by 
\begin{equation}
A=\hbar \omega _{c}a^{\dagger j_{\downarrow }}a^{j_{\uparrow }},\qquad
A^{\dagger }=\hbar \omega _{c}a^{\dagger j_{\uparrow }}a^{j_{\downarrow }},
\label{GenerAnnih}
\end{equation}%
with $j_{\uparrow }$ and $j_{\downarrow }$ being integers such that $%
j_{\uparrow }>j_{\downarrow }$. The monolayer graphene is given by $%
j_{\uparrow }=1$, $j_{\downarrow }=0$, and the bilayer graphene by\ $%
j_{\uparrow }=2$, $j_{\downarrow }=0$. The physical meaning of the indices
is discussed after the energy spectrum is constructed.

We study the energy spectrum of the Hamiltonian $H^{+}=\{Q_{+},Q_{+}\}$ with
(\ref{SuperQ}) and (\ref{GenerAnnih}). Although the Hamiltonian is written
in terms of $A$ and $A^{\dag }$, the basic physical variable is the
momentum, or equivalently, $a$ and $a^{\dag }$. Thus, we consider the state 
\begin{equation}
|n\rangle =\frac{1}{\sqrt{n!}}(a^{\dag })^{n}|0\rangle .
\end{equation}%
We make an analysis at the K point. The zero-energy up-spin states are given
by the condition $A|n\rangle =0$. They are $|0\rangle $, $|1\rangle $, $%
\cdots $, $|j_{\uparrow }-1\rangle $, which are degenerate in $|\psi
_{0}^{+\uparrow }\rangle $. On the other hand the zero-energy down-spin
states are determined by requiring $A^{\dag }|n\rangle =0$. If $%
j_{\downarrow }=0$, there are no zero-energy states, and the supermultiplet
is given by (\ref{SuperMultiA}) [Fig.\ref{FigSuperPair}(a)]. If $%
j_{\downarrow }\neq 0$, they are $|0\rangle $, $|1\rangle $, $\cdots $, $%
|j_{\downarrow }-1\rangle $, which are degenerate in $|\psi
_{0}^{+\downarrow }\rangle $. Consequently, there exists the ($j_{\uparrow
}+j_{\downarrow }$)-fold degeneracy in the zero-energy state for electrons
at K point, corresponding to $j_{\uparrow }$ up-spin electrons and $%
j_{\downarrow }$ down-spin electrons [Fig.\ref{FigSuperPair}(b)]. The
similar analysis is made also at the K' point.

We apply the above analysis to the "Dirac" Hamiltonian (\ref{DiracHamilC}).
The\ eigenvalues are easily calculated by evaluating $A^{\dagger
}A\left\vert n\right\rangle $ and $AA^{\dagger }\left\vert n\right\rangle $.
For the bosonic sector, we derive 
\begin{equation}
\mathcal{E}_{0}^{+\uparrow }=\mathcal{E}_{0}^{-\downarrow }=\pm mv_{\text{F}%
}^{2}
\end{equation}
and 
\begin{equation}
\mathcal{E}_{n}^{+\uparrow }=\mathcal{E}_{n}^{-\downarrow }=\varepsilon
^{+\uparrow }(j_{\uparrow }+n-1)
\end{equation}
for $n\geq 1$, where%
\begin{equation}
\varepsilon ^{+\uparrow }(n)=\pm \hbar \omega _{c}\sqrt{\frac{n!\left(
n-j_{\uparrow }+j_{\downarrow }\right) !}{\left\{ \left( n-j_{\uparrow
}\right) !\right\} ^{2}}+\frac{m^{2}v_{\text{F}}^{4}}{\hbar ^{2}\omega
_{c}^{2}}}.
\end{equation}%
For the fermionic sector, in the case of $j_{\downarrow }=0$\ we find 
\begin{equation}
\mathcal{E}_{n}^{+\downarrow }=\mathcal{E}_{n}^{-\uparrow }=\varepsilon
^{+\downarrow }(j_{\downarrow }+n)
\end{equation}
for $n\geq 0$ [Fig.\ref{FigSuperPair}(a)]; in the case of $j_{\downarrow
}\neq 0$ we find 
\begin{equation}
\mathcal{E}_{0}^{+\downarrow }=\mathcal{E}_{0}^{-\uparrow }=\pm mv_{\text{F}%
}^{2}
\end{equation}
and 
\begin{equation}
\mathcal{E}_{n}^{+\downarrow }=\mathcal{E}_{n}^{-\uparrow }=\varepsilon
^{+\downarrow }(j_{\downarrow }+n-1)
\end{equation}
for $n\geq 1$ [Fig.\ref{FigSuperPair}(b)], where 
\begin{equation}
\varepsilon ^{+\downarrow }(n)=\pm \hbar \omega _{c}\sqrt{\frac{n!\left(
n+j_{\uparrow }-j_{\downarrow }\right) !}{\left\{ \left( n-j_{\downarrow
}\right) !\right\} ^{2}}+\frac{m^{2}v_{\text{F}}^{4}}{\hbar ^{2}\omega
_{c}^{2}}}.
\end{equation}%
In the case of $m=0$, the zero energy states are $4(j_{\uparrow
}+j_{\downarrow })$-fold degenerated and all other states are $4$-fold
degenerated. This energy spectrum implies that the Hall conductivity is
quantized as in (\ref{HallConduA}).

The bilayer graphene corresponds to $j_{\uparrow }=2$, $j_{\downarrow }=0$.
The energy spectrum is obtained as 
\begin{equation}
\mathcal{E}_{0}^{+\uparrow }=\mathcal{E}_{0}^{-\downarrow }=\pm mv_{\text{F}%
}^{2}
\end{equation}
and\beginABC%
\begin{equation}
\mathcal{E}_{n}^{+\uparrow }=\mathcal{E}_{n}^{-\downarrow }=\pm \hbar
\left\vert \omega _{c}\right\vert \sqrt{n\left( n+1\right) +m^{2}v_{\text{F}%
}^{4}/\hbar ^{2}\omega _{c}^{2}},
\end{equation}%
for $n\geq 1$ and%
\begin{equation}
\mathcal{E}_{n}^{+\downarrow }=\mathcal{E}_{n}^{-\uparrow }=\pm \hbar
\left\vert \omega _{c}\right\vert \sqrt{\left( n+2\right) \left( n+1\right)
+m^{2}v_{\text{F}}^{4}/\hbar ^{2}\omega _{c}^{2}}
\end{equation}%
\endABC for $n\geq 0$, as agrees with the previous result\cite{Nov3,MacCann}%
. We expect the emergence of a Hall plateau at $\nu =0$ due to the
electron-hole splitting also in the bilayer graphene.

We now discuss the physical meaning of the indices $j_{\uparrow }$ and $%
j_{\downarrow }$ in (\ref{GenerAnnih}) by setting $m=0$. We derive the
energy dispersion relation in the zero field ($B=0$). Setting $P_{i}=p_{i}$
and parametrizing $\mathbf{p}=\left( p\cos \phi ,p\sin \phi \right) $, we
obtain%
\begin{equation}
A=E\left( p\right) e^{i\left( j_{\uparrow }-j_{\downarrow }\right) \phi
},\quad A^{\dagger }=E\left( p\right) e^{-i\left( j_{\uparrow
}-j_{\downarrow }\right) \phi },
\end{equation}%
with 
\begin{equation}
E\left( p\right) =\hbar \omega _{c}\left( \frac{l_{B}p}{\sqrt{2}\hbar }%
\right) ^{j_{\uparrow }+j_{\downarrow }}.
\end{equation}
The eigenvalues of the "Dirac" Hamiltonian (\ref{DiracHamilC}) are $\pm
E\left( p\right) $ with the eigenstates 
\begin{equation}
\left\vert \pm \right\rangle =\frac{1}{\sqrt{2}}\left( 
\begin{array}{c}
1 \\ 
\pm e^{i\left( j_{\uparrow }-j_{\downarrow }\right) \phi }%
\end{array}%
\right)
\end{equation}
at the K point. Hence, the energy dispersion relation is given by $E\left(
p\right) $. The Berry phase is defined as a loop integration 
\begin{equation}
\Gamma _{\text{B}}=\oint \sum_{k}C_{k}dx_{k}
\end{equation}
of the connection field 
\begin{equation}
C_{k}=\left\langle n\right\vert i\frac{\partial }{\partial x_{k}}\left\vert
n\right\rangle .
\end{equation}
It becomes 
\begin{equation}
\Gamma _{\text{B}}=\pi \left( j_{\uparrow }-j_{\downarrow }\right)
\end{equation}
for the eigenstates $\left\vert \pm \right\rangle $. Thus the indices $%
j_{\uparrow }$ and $j_{\downarrow }$ are fixed by the dispersion relation
and the Berry phase in the zero field.

We comment on the Witten index and the Atiyah-Singer index theorem. We focus
on the K point. The Witten index\cite{Witten} is given by%
\begin{equation}
\Delta _{\text{W}}=\dim \left[ \ker H^{+\uparrow }\right] -\dim \left[ \ker
H^{+\downarrow }\right] =j_{\uparrow }-j_{\downarrow },
\end{equation}%
where $j_{\uparrow }\equiv \dim \left[ \ker H^{+\uparrow }\right] $ and $%
j_{\downarrow }\equiv \dim \left[ \ker H^{+\downarrow }\right] $ are the
numbers of zero-energy states of $H^{+\uparrow }$ and $H^{+\downarrow }$,
respectively. We have explicitly shown in the present model that the Witten
index defined in the nonzero field ($B\neq 0$) is equal to the Berry phase
defined in the zero field ($B=0$), $\Gamma _{\text{B}}=\pi \Delta _{\text{W}%
} $. This is a general property, as we briefly sketch. The Witten index is
equal to the Fredholm index\cite{Thaller92}%
\begin{equation}
\Delta _{\text{F}}=\dim \left[ \ker A\right] -\dim \left[ \ker A^{\dag }%
\right] ,
\end{equation}%
since 
\begin{equation}
\ker H^{+\uparrow }=\ker A^{\dagger }A=\ker A,
\end{equation}%
and 
\begin{equation}
\ker H^{+\downarrow }=\ker AA^{\dagger }=\ker A^{\dagger }.
\end{equation}
Now, according to the Atiyah-Singer index theorem, $\Delta _{\text{F}}$ is
invariant as $B\rightarrow 0$, where $\Delta _{\text{F}}$ becomes the chiral
anomaly\cite{Thaller92}. The chiral anomaly is given by the Berry phase.
Consequently it follows that $\Gamma _{\text{B}}=\pi \Delta _{\text{W}}$.

The Witten index, $j_{\uparrow }-j_{\downarrow }$, and the degeneracy of the
zero-energy state, $j_{\uparrow }+j_{\downarrow }$, are different objects.
Nevertheless, they are identical in graphene since $j_{\downarrow }=0$.
Thus, by combining the contributions from the K and K' points, the
degeneracy $4j_{\uparrow }=4$ in monolayer graphene can be related to the
Berry phase, as pointed out in Refs.\cite{Nov2,Zhang}.

We have presented a unified description of the QHE in graphene based on the
SUSY quantum mechanics. The key observation is that the Zeeman splitting is
exactly as larage as the Landau level separation at the K and K' points in
graphene. It is remarkable that its consequence is the exact spin SU(2)
symmetry in each Landau level, however large the magnetic field is. Indeed,
an arbitral linear combination of the up-spin and down-spin states belongs
to the same Landau level. Furthermore, the electron has the pseudospin in
addition to the spin. It leads to the SU(4) symmetry together with SU(4)
skyrmion excitations\cite{ZFE}, as in the bilayer system of the conventional
semiconductor QHE\cite{BookEzawa}.

Interactions between electrons are not taken into account in this paper,
except for a possible mass term which would arise from Coulomb interactions.
It removes the $4$-fold degeneracy in the zero-energy state of monolayer
graphene, as explains the emergence of a Hall plateau at $\nu =0$ in recent
experimental data\cite{Zhang06L}. However, it does not remove the $4$-fold
degeneracy in higher Landau levels. With respect to this degeneracy, there
are senarios\cite{Nomura,Fisher} that the degeneracy is removed by Coulomb
interaction. In particular, it is pointed out\cite{Fisher} that the
pseudospin SU(2) symmetry is broken explicitity down to U(1)$\times $Z$_{%
\text{2}}$. If this is the case, the total symmetry must be [SU(2)]$_{\text{%
spin}}\times $[U(1)$\times $Z$_{\text{2}}$]$_{\text{pseudospin}}$, since the
spin SU(2) symmetry is exact as we have emphasized. We wish to make a
detailed analysis of a fine structure of Hall plateaux based on our SUSY
formalism in a subsequent work.

The author is grateful to Professors T. Ando, Y. Iye and Z.F. Ezawa for
fruitful discussions on the subject.

\end{document}